\documentclass[letter]{aa}
\usepackage{graphicx}
\usepackage[varg]{txfonts}
\usepackage{natbib}
\newdimen\captwidth
\newdimen\figwidth

\newcommand{\rd}{\mathrm{d}}
\newcommand{\excs}{\extracolsep{\fill}}

\begin{document}

\title{Orbital clustering of distant Kuiper Belt Objects by hypothetical Planet 9. Secular or resonant ?}
\author{H. Beust\inst{1,2}}
\institute{Universit\'e Grenoble Alpes, IPAG, CS 40700, F-38058 Grenoble 
Cedex 9, France \and
CNRS, IPAG, CS 40700, F-38058 Grenoble 
Cedex 9, France}
\date{Received ....; Accepted....}
\offprints{H. Beust}
\mail{Herve.Beust@univ-grenoble-alpes.fr}
\titlerunning{Orbital clustering of KBO's by Planet 9}
\authorrunning{H. Beust}
\abstract{Statistical analysis of the orbits of distant Kuiper Belt Objects (KBOs) have led to suggest that an additional planet should reside in the Solar System. According to recent models, the secular action of this body should cause orbital alignment of the KBOs.}
{It was recently claimed that the KBOs concerned by this dynamics are presumably trapped in mean motion resonances with the suspected planet. I reinvestigate here the secular model underlying this idea.}
{The original analysis was done expanding and truncating the secular Hamiltonian. I show that this is inappropriate here, as the series expansion is not convergent. I present a study based on numerical computation of the Hamiltonian with no expansion.}  
{I show in phase-space diagrams the existence of apsidally anti-aligned, high eccentricity libration islands that were not present in the original modelling, but that match numerical simulations. These island were claimed to correspond to bodies trapped in mean-motion resonances with the hypothetical planet, and match the characteristics of the distant KBOs observed .} 
{My main result is that regular secular dynamics can account for the anti-aligned particles itself as well as mean-motion resonances. I also perform a semi-analytical study of resonant motion and show that some resonance are actually capable of producing the same libration islands. I discuss then the relative importance of both mechanisms.}  
\maketitle
\section{Introduction}
The growing statistics about distant Kuiper Belt Objects (KBOs) recently revived interest into the possible presence of an additional distant planet (hereafter termed Planet 9 or P9) in the Solar System. This came out after the discovery of 2012~VP$_{113}$, a large KBO with orbital parameters similar to those of Sedna \citep{tru14}. As noted by \citet{fue14}, all such objects with large perihelia and eccentricities have arguments of perihelia $\omega$ concentrating around $0$. Even if the statistics is poor, such a distribution is unlikely, as orbital precession induced by the giant planets is expected to quickly randomize $\omega$ values. This led \citet{fue14} and \citet{tru14} to suggest that the perturbing action of a distant Super-Earth sized planet could help maintaining this apsidal clustering. 

This issue was recently investigated into more details by \citet{bat16} (hereafter B16). They first note that the distant KBOs not only gather around $\omega=318\degr\pm8\degr$, but that the same applies to their longitudes of ascending nodes $\Omega$ which satisfy $\Omega=113\degr\pm13\degr$. As a consequence the orbits of all bodies concerned are physically roughly aligned. Based on the idea that this orbital confinement is due to the secular action of the suspected P9, they develop a secular dynamical model to constrain its parameters. They come to the conclusions that the planet that best reproduces the observational data has mass $m'=10\,\mbox{M}_\oplus$, semi-major axis $a'=700\,$au and eccentricity $e'=0.6$.

Further constraints on this planet are hard to derive. Based on the analysis of residuals in the orbital motion of Saturn recorded by the Cassini spacecraft in the last decade, \citet{fie16} show recently that the residuals are better explained by P9 if we give it a current true anomaly around $\sim 120\degr$. This can indicate a preferred region to try to detect it. \citet{cow16} claim that if it is large enough, P9 could be detected as a 30\,mJy at 1\,mm wavelength by existing cosmology experiments.

The purpose of this paper is to reinvestigate B16's secular model and conclusions. More specifically, B16 first develop a semi-analytical dynamical model of distant particles as secularly perturbed by the giant planets and the hypothetical P9, and show that if it is eccentric enough, P9 can actually confine the orbits in an apsidally aligned configuration with respect to it. Then they move to numerical simulations. They recover the apsidally resonant islands noted in their semi-analytical study, but note that the particles trapped there do not have high enough eccentricities to account for the population of distant KBOs under consideration. Conversely, they notice the presence of new, high eccentricity and anti-aligned libration islands in their numerical work that did not appear in the semi-analytical work. They claim that these anti-aligned particles are presumably trapped in mean-motion resonances (hereafter MMRs) with P9, as resonant dynamics is not taken into account in the secular theory. They show that some particles in their numerical study do exhibit resonant trapping. 

This issue was further reinvestigated by \citet{mal16}, who found that the orbital periods of the main distant KBOs present commensurabilities that could indicate resonant configurations with P9. According to \citet{mal16}, Sedna is putatively trapped in 3:2 MMR with P9.

I reinvestigate here the latter hypothesis. I show that the semi-analytical analysis of B16 is inappropriate to the present case. The reason is the assumed expansion of the secular Hamiltonian that is not convergent. I thus perform a full numerical computation of the secular Hamiltonian, and show that anti-aligned, high eccentricity libration islands that were not present in the simplified analysis appear in the non-resonant phase-space maps. Hence invoking resonant trapping might not be necessary. I subsequently investigate the resonant dynamics in a similar semi-analytical manner, and show that various MMRs can actually also generate high eccentricity, anti-aligned librating particles as well, as suggested by B16. I discuss and compare the relevance of both mechanisms.
\section{Non-resonant secular dynamics}
\begin{figure*}
\makebox[\textwidth]{
\includegraphics[width=0.33\textwidth]{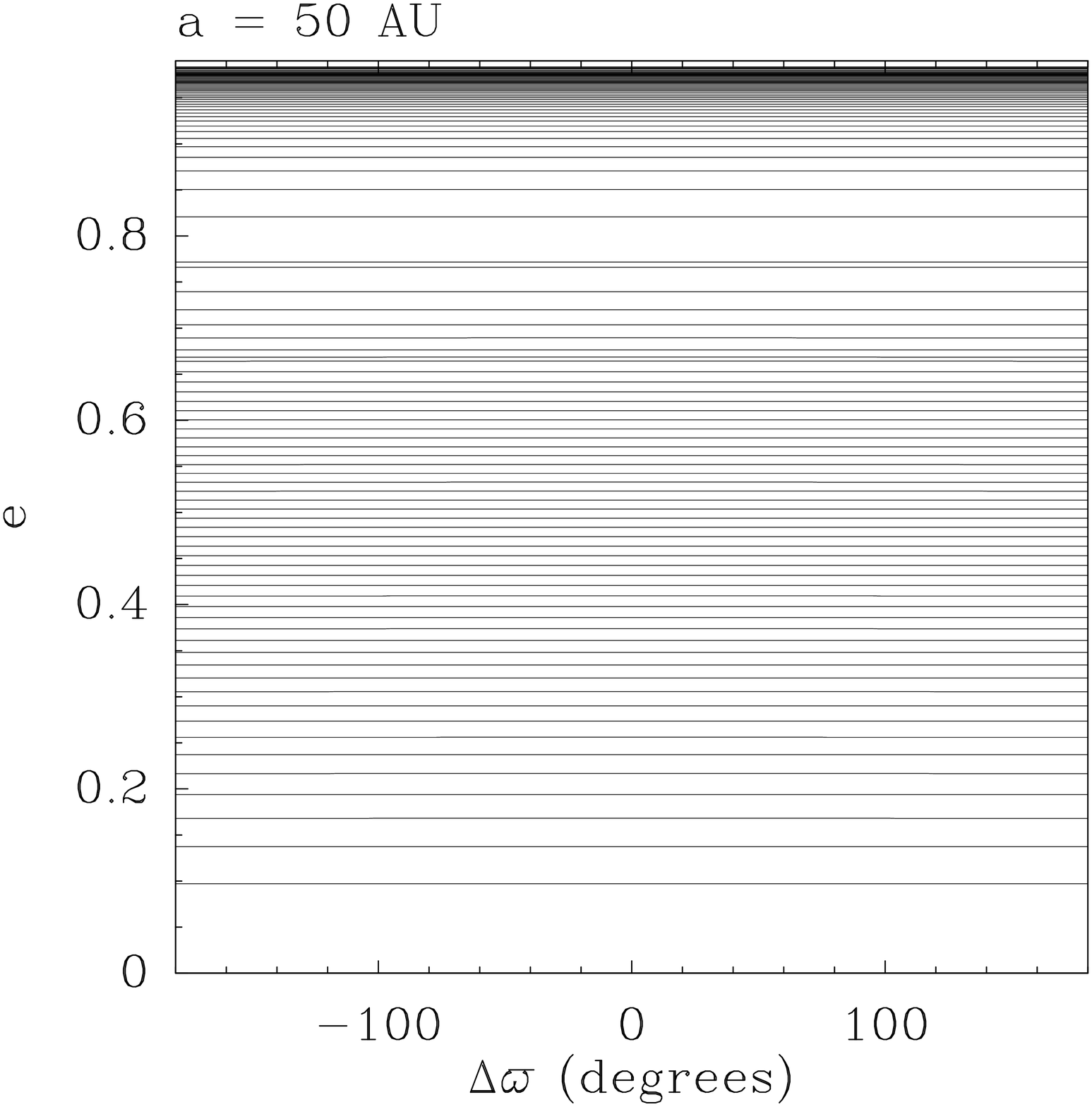} \hfil
\includegraphics[width=0.33\textwidth]{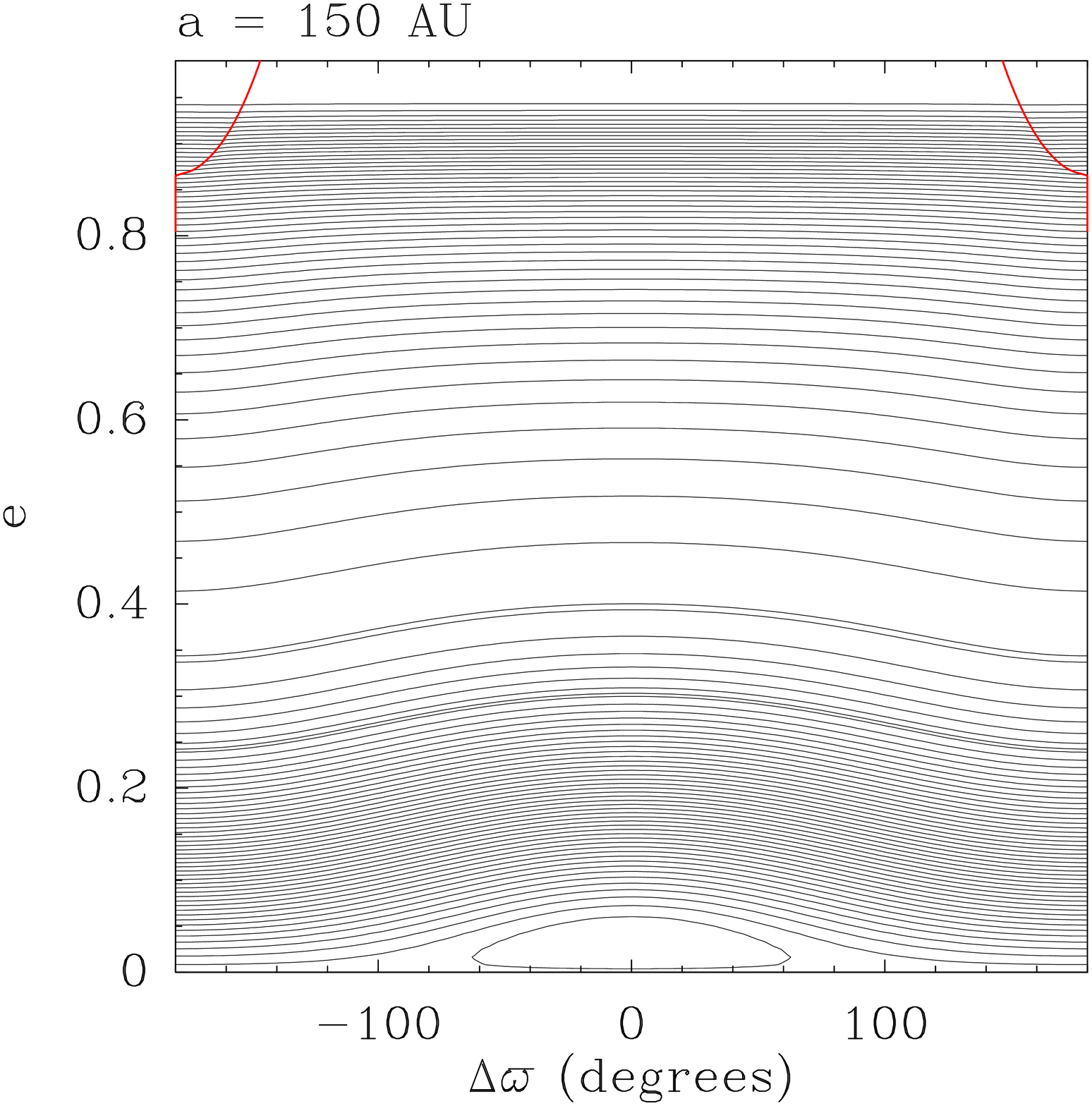} \hfil
\includegraphics[width=0.33\textwidth]{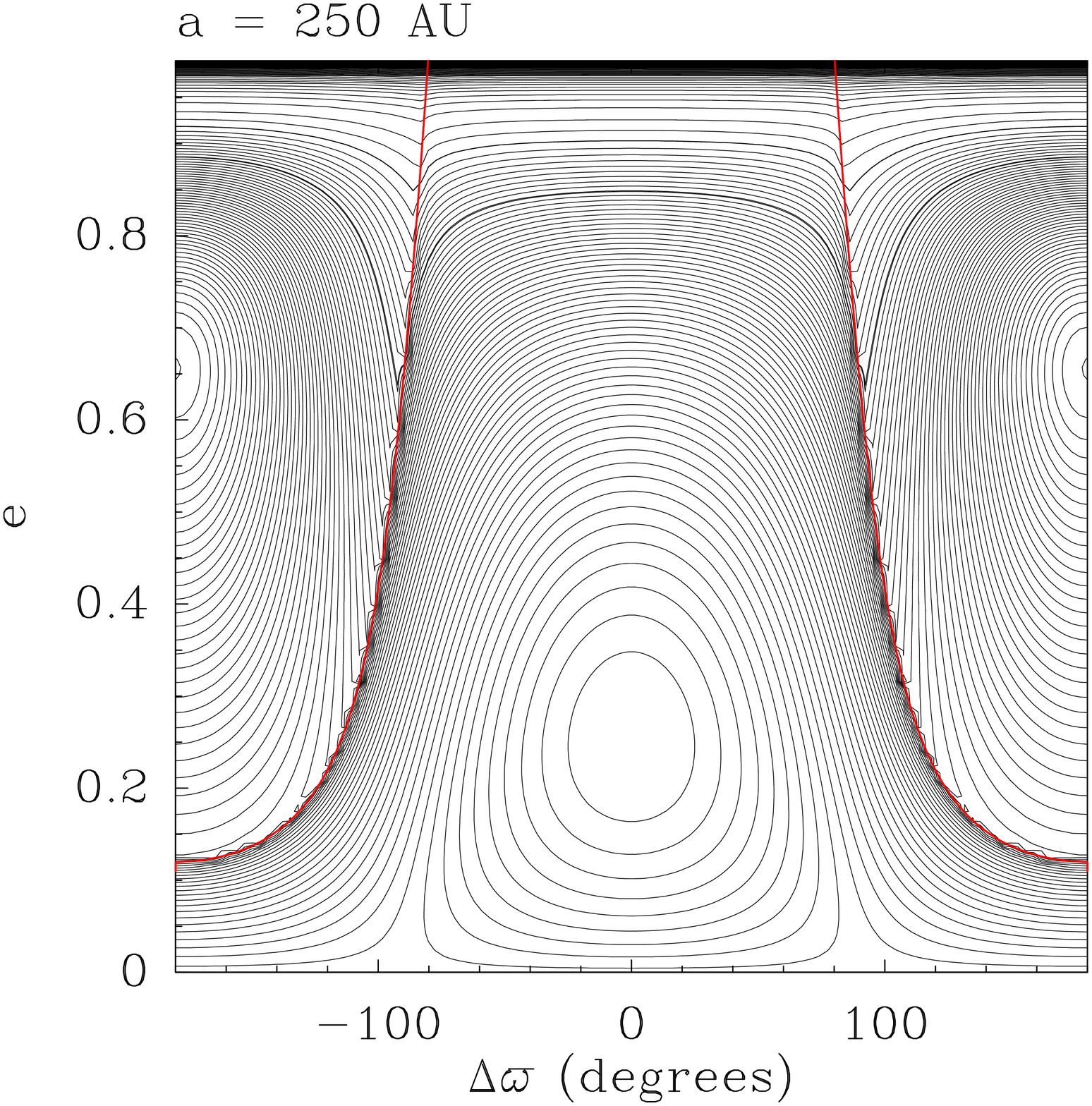}}
\vspace*{-\jot}\\
\makebox[\textwidth]{
\includegraphics[width=0.33\textwidth]{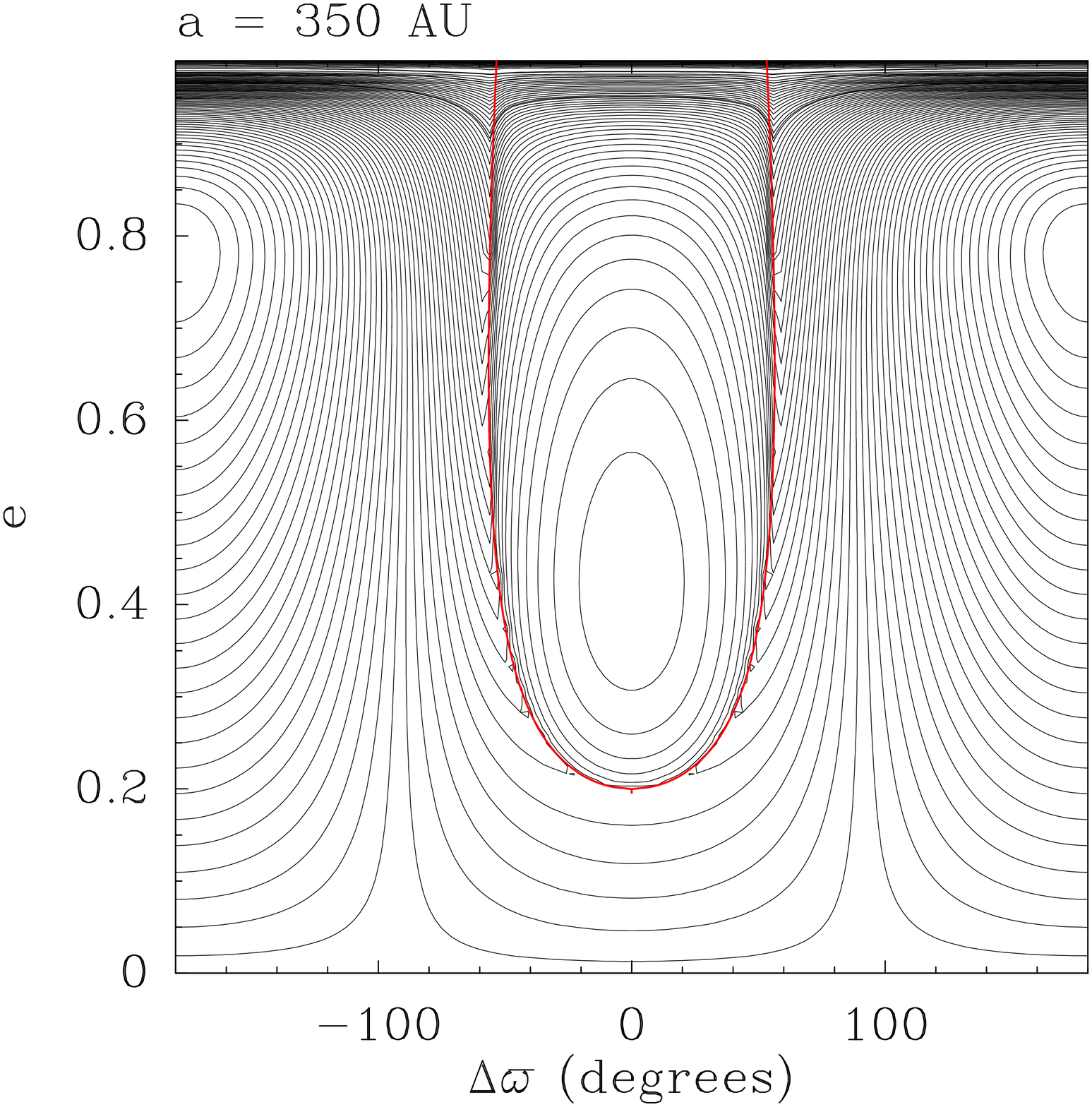} \hfil
\includegraphics[width=0.33\textwidth]{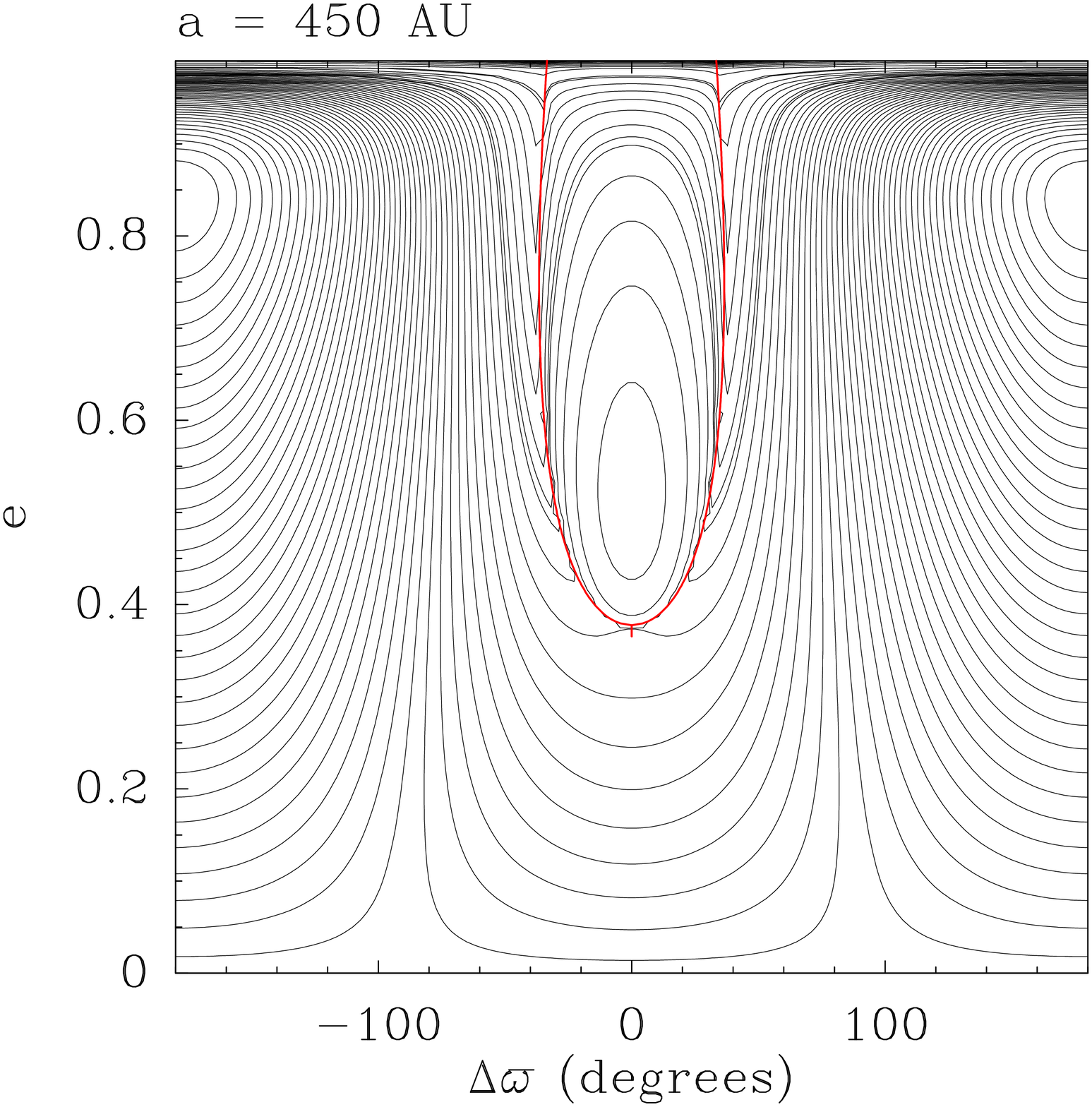} \hfil
\includegraphics[width=0.33\textwidth]{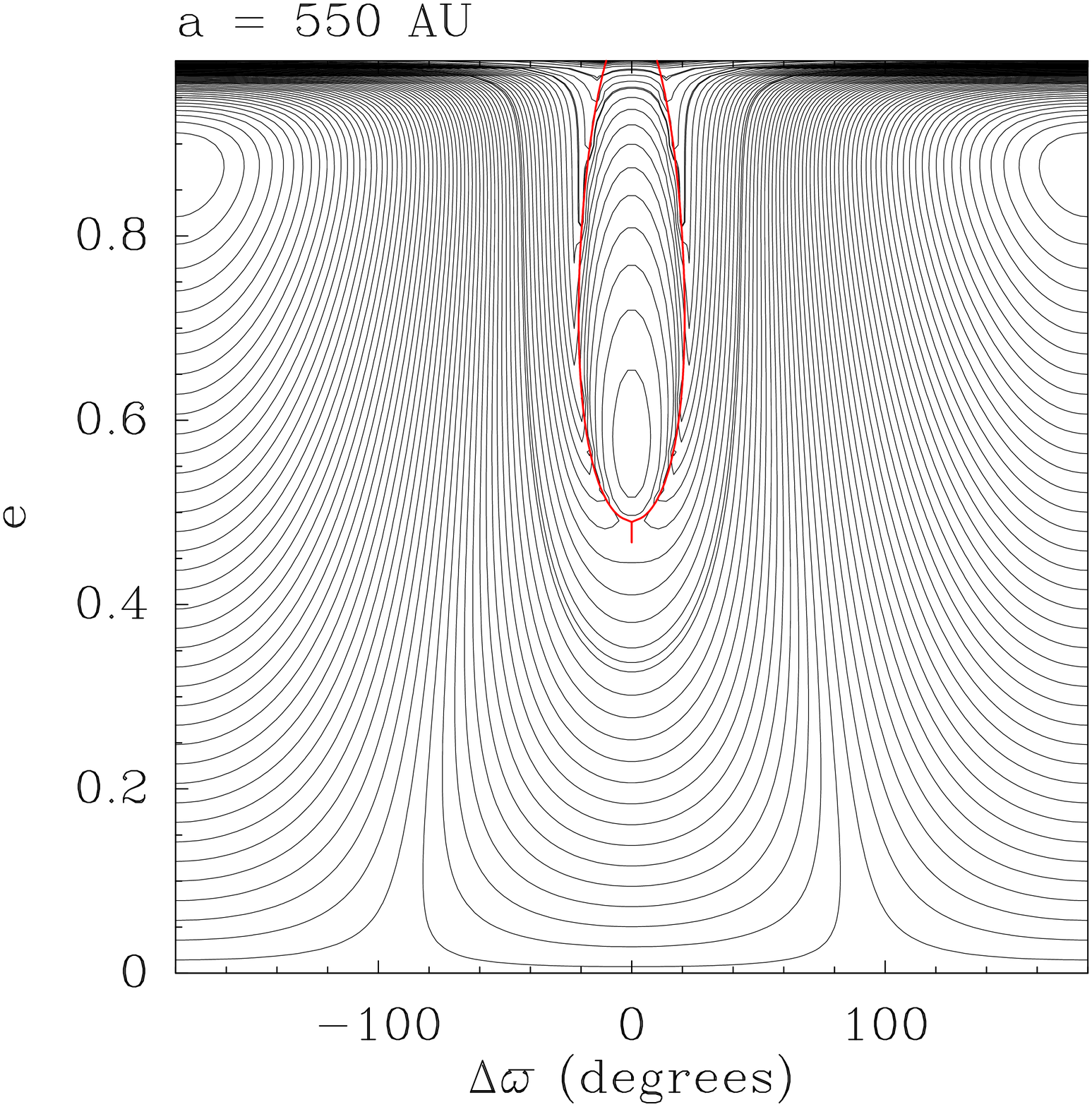}}
\caption[]{Phase portraits of untruncated, secular averaged Hamiltonian (\ref{htot}) in $(\Delta\varpi,e)$ space, computed in the same conditions as B16, i.e., assuming parameters $m'=10\,M_\oplus$, $a'=700\,$au and $e'=0.6$ for P9. Plots are drawn for various semi-major axis values listed on top of the panels. On each plot, the red curve separates regions where both orbits cross from regions where they do not. In the $a=50\,$au case, the orbits never cross; for $a=150\,$au and $a=250\,$au, the orbits cross in the regions located above the red curves around $\Delta\varpi=180\degr$; for $a=350\,$au, $a=450\,$au and $a=550\,$au, the orbits cross in most configurations, except above the red curves centered around $\Delta\varpi=0\degr$ where both orbits remain nested.}
\label{hsec_map}
\end{figure*}
Consider a massless particle test orbiting the Sun, perturbed by Planet~9.
In the following, all applications are done with the following parameters for P9 : $m'=10\,\mbox{M}_\oplus$, $a'=700\,$au and $e'=0.6$, as this appears to be the best model quoted by B16. As long as MMRs are not concerned, the long-term behaviour of the particle's orbit is well described with the secular Hamiltonian, which is obtained taking the time average of the instantaneous interaction Hamiltonian over both orbits independently:
\begin{eqnarray}
\lefteqn{H_\mathrm{sec} \; = \; -\frac{Gm'}{4\pi^2}\int_0^{2\pi}\!\!\int_0^{2\pi}
\left(\frac{1}{|\vec{r}-\vec{r'}|}-\frac{\vec{r}\cdot
\vec{r'}}{r'^3}\right)\,\rd l\,\rd l'}&&\nonumber\\
\lefteqn{-\frac{1}{4}\frac{GM}{a}\left(1-e^2\right)^{-3/2}
\sum_{i=1}^4\frac{m_ia_i^2}{Ma^2}
+\dot{\varpi}'\sqrt{GMa}\left(1-\sqrt{1-e^2}\right)\;,}&&
\label{htot}
\end{eqnarray} 
where $M$ is the mass of the Sun; $\vec{r}$ and $\vec{r'}$ are the instantaneous heliocentric radius vectors of the particle and of P9, respectively; $l$ and $l'$ are their mean anomalies; $a$ is the particle's semi-major axis and $e$ its eccentricity. See Appendix~\ref{hsecapp} for the derivation of this expression. The last two terms account for the perturbing action of the known giant planets on the particle and on P9. Here in the second term, the sum extends over the four giant planets, each of them having mass $m_i$ and semi-major axis $a_i$; $\dot{\varpi}'$ is the precession rate of P9's perihelion caused by the same planets.

This Hamiltonian cannot be expressed in closed form though. However, taking advantage from the fact that the particle's orbit lies inside P9's one, the disturbing function can be expanded in ascending powers of $a/a'$. The integral appearing in Eq.~(\ref{htot}) can be written as:
\begin{equation}
-\frac{Gm'}{4\pi^2}\int_0^{2\pi}\!\!\int_0^{2\pi}
\left(\frac{1}{|\vec{r}-\vec{r'}|}-\frac{\vec{r}\cdot
\vec{r'}}{r'^3}\right)\,\rd l\,\rd l'=-\frac{Gm'}{a'}\sum_{k=2}^{+\infty}h_k\;,
\end{equation}
where the $h_k$'s are dimensionless coefficients that can be expressed in closed form \citep{las10}. Each $h_k$ is proportional to $(a/a')^k$. For coplanar orbits, they also are functions of $e$, $e'$ and of $\Delta\varpi=\varpi'-\varpi$ only, where $\varpi$ and $\varpi'$ are the longitudes of periastron of the particle and of P9 respectively. Whenever $a<a'$, the $h_k$'s are generally assumed to decrease with increasing order $k$, so that the expansion can be truncated to some finite order $n$. The Hamiltonian given in Eq.~(4) of B16 corresponds to a truncation at $n=3$ order (octupole approximation).

Two potential problems may arise with this approach, if one wants 
the truncated $H_\mathrm{sec}$ to accurately represent the actual dynamics of the particle. First, to ensure the validity of the secular model, the particle must not be trapped in any MMR, and remain protected from close encounters. \citet{beu14} showed however that even in the case of crossing orbits, the secular theory provides a relevant description of the dynamics as long as a close encounter does not occur. Second, one must ensure that the series expansion of the secular Hamiltonian is convergent and that the truncation order is large enough to allow the truncation to accurately approximate to the full sum. 

\begin{table}
\caption[]{Numerical values of $h_k$'s for various values of orbital parameters corresponding to configurations tested by B16. In all cases, I assume $\Delta\varpi=0$, $a'=700\,$au and $e'=0.6$.}
\label{hkvalues}
\begin{tabular*}{\columnwidth}{@{\excs}lllll}
\hline\noalign{\smallskip}
$a$ & 450 au & 450 au & 150 au & 150 au\\
$e$ & 0.1 & 0.6 & 0.1 & 0.6\\
%$\Delta\varpi$ & 0 & 0 & 0 & 0\\
\noalign{\smallskip}\hline\noalign{\smallskip}
$h_2$ & $-0.046$ & $0.311$ & $2.27\times10^{-2}$ & $3.45\times10^{-2}$\\
$h_3$ & $-0.046$ & $-0.348$ & $-1.70\times10^{-3}$ & $-1.29\times10^{-2}$\\
$h_4$ & $0.189$ & $0.690$ & $2.33\times10^{-3}$ & $8.52\times10^{-3}$\\
$h_5$ & $-0.105$ & $-1.28$ & $-4.32\times10^{-4}$ & $-5.26\times10^{-3}$\\
$h_{10}$ & $1.15$ & $55.0$ & $1.94\times10^{-5}$ & $9.32\times10^{-4}$\\
$h_{20}$ & $110.6$ & $2.55\times10^5$ & $3.17\times10^{-8}$ &
$7.32\times10^{-5}$\\
$h_{50}$ & $7.35\times10^8$ & $1.33\times10^{17}$ & $1.02\times10^{-15}$ &
$1.86\times10^{-7}$\\
\noalign{\smallskip}\hline
\end{tabular*}
\end{table}
Convergence of the series expansion is usually ensured for small enough $a/a'$ ratio. In the configurations described in B16 with orbits that sometimes cross, this is far from being obvious. Table~\ref{hkvalues} lists various numerical values of $h_k$'s computed for different configurations of the particle's orbit. For configurations with $a=150$\,au, the particle's orbit lies well inside that of the planet, whilst when $a=450\,$au, both orbits cleary cross. In the former case, the $h_k$'s rapidly decrease, ensuring convergence of the expansion, but in the latter case it is obviously divergent.

An alternate way to avoid the expansion is to compute numerically the double integral appearing in Eq.~(\ref{htot}). The integral is computed using Gauss-Legendre numerical quadrature with $70\times70$ points, and with special care of the regions where the orbits cross. This technique was already applied to the semi-analytic study of Fomalhaut~b's dynamics in \citet{beu14}. The result is shown in Fig.~\ref{hsec_map}. This figure was built assuming the same input conditions as Fig.~3 from B16, except that the latter was computed assuming octupolar approximation. On each plot, the red curve marks the separation between regions where both orbits actually cross and regions where they do not. 

Although the general shape of the various phase portraits agree in both cases, striking differences nevertheless appear. In all situations, libration islands around $\Delta\varpi=0\degr$ (i.e., apsidal alignment) appear, but for large $a$, they are systematically narrower and centered around a secular equilibrium located at lower eccentricity than in the octupolar approximation. Moreover, for $a\geq 250\,$au, new libration islands appear around $\Delta\varpi=180\degr$ (i.e., antialignement) at high eccentricity. These are not present in Fig.~3 from B16. 

My phase portraits appear to much more closely match the results of B16's numerical exploration (their Fig.~4) than their original maps. The apsidally aligned libration islands now appear at the right eccentricity with corresponding widths. Moreover, the anti-aligned libration islands reported by B16 in their numerical exploration exactly match those I get in my phase portaits. These anti-aligned librating particles are of prime importance in B16's discussion, as they suggest that the apsidally confined distant KBOs could be trapped in this dynamical state. 

B16 claimed that the particles exhibiting this behaviour in their numerical exploration were actually trapped in a MMR with the perturbing planet. Resonant dynamics in indeed not described by the secular theory, irrespective of whether the Hamiltonian is truncated or not. B16 show indeed some particles out of their simulation that appear to be trapped in MMRs, at least temporarily. Whilst there is no doubt about the reality of resonant behaviours, I claim here that secular dynamics is also able to generate the anti-aligned, high eccentricity librating particles they find in their numerical exploration as well.
\section{Resonant dynamics} 
\begin{figure*}
\makebox[\textwidth]{
\includegraphics[width=0.33\textwidth]{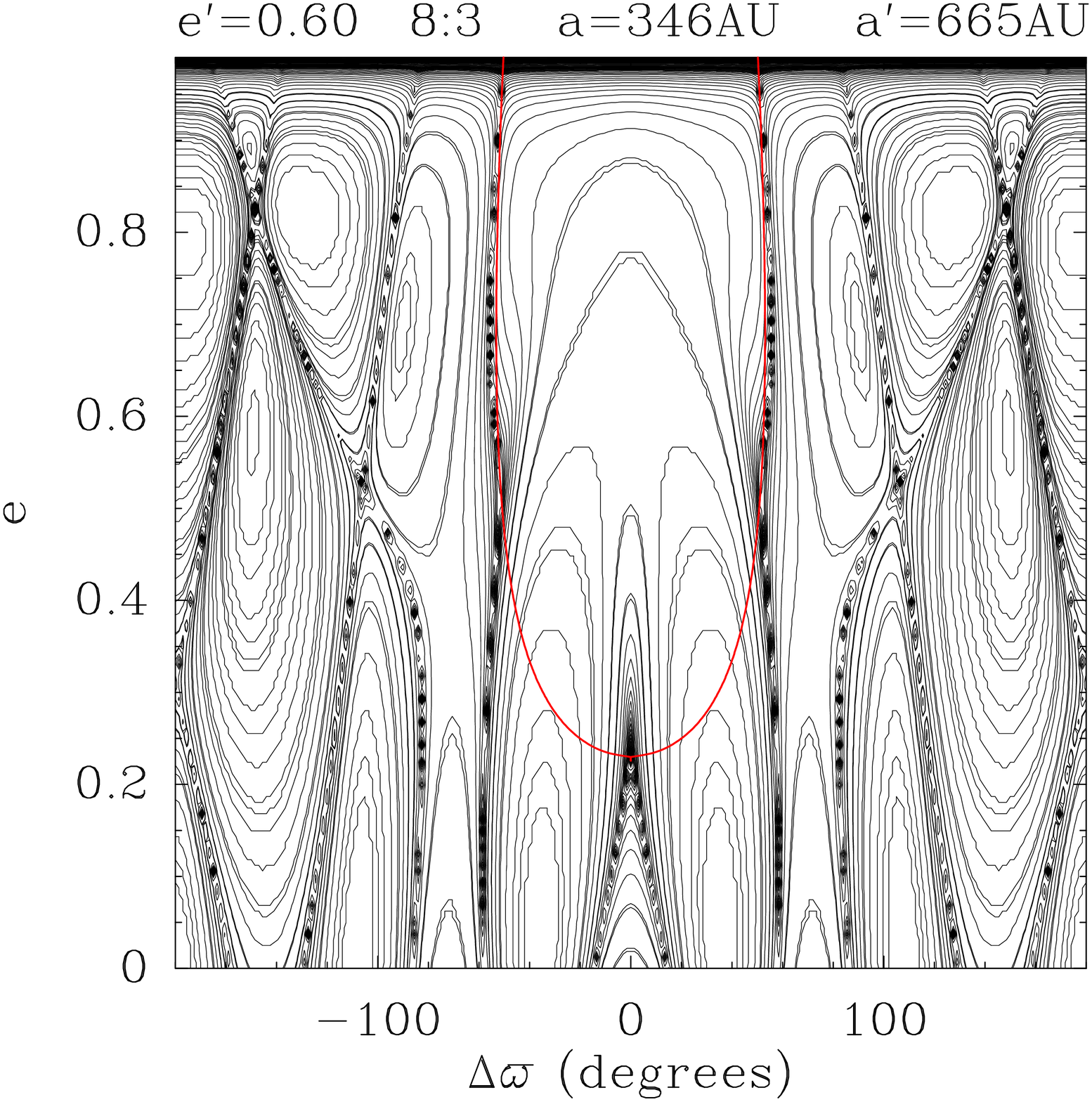} \hfil
\includegraphics[width=0.33\textwidth]{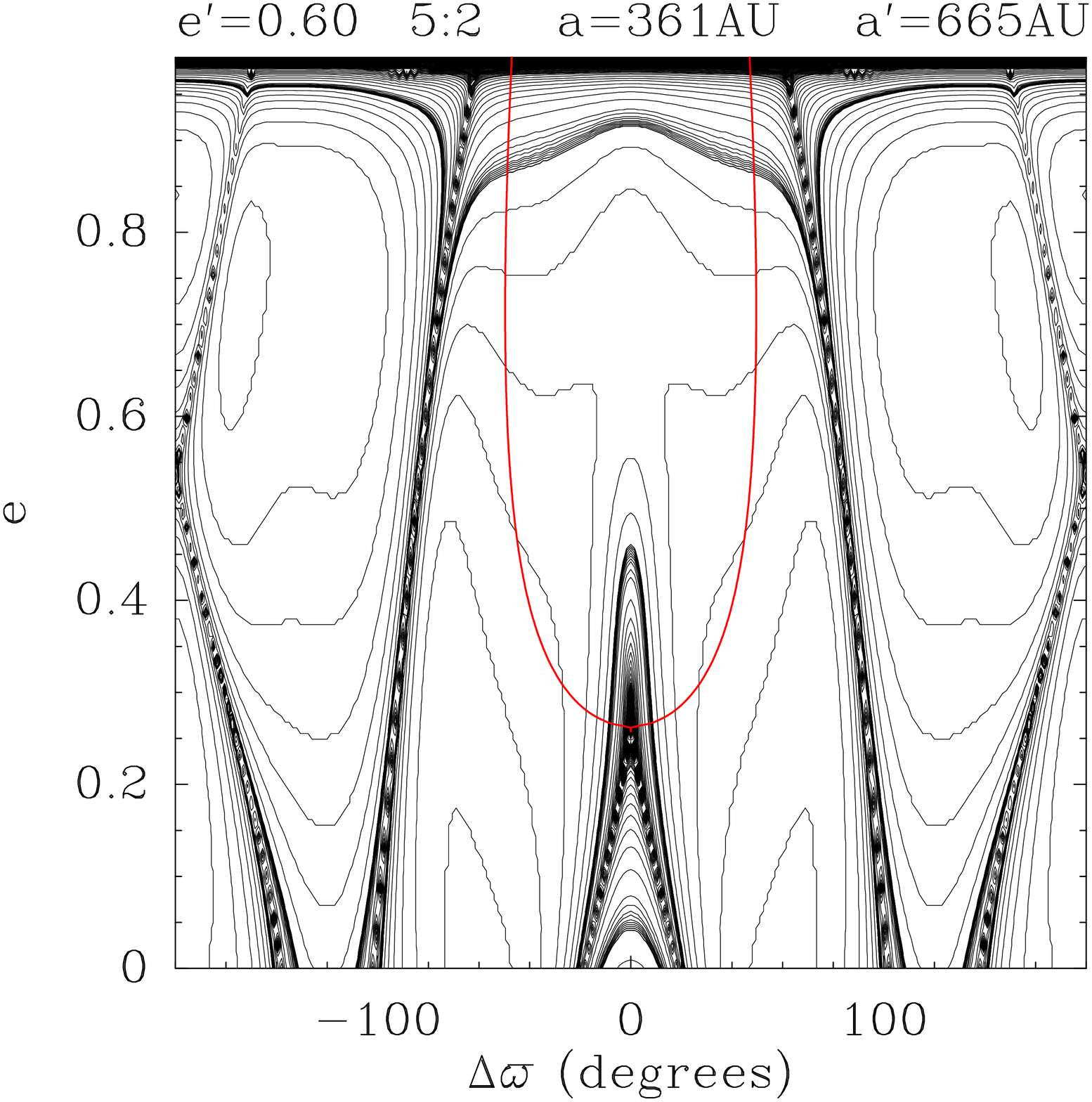} \hfil
\includegraphics[width=0.33\textwidth]{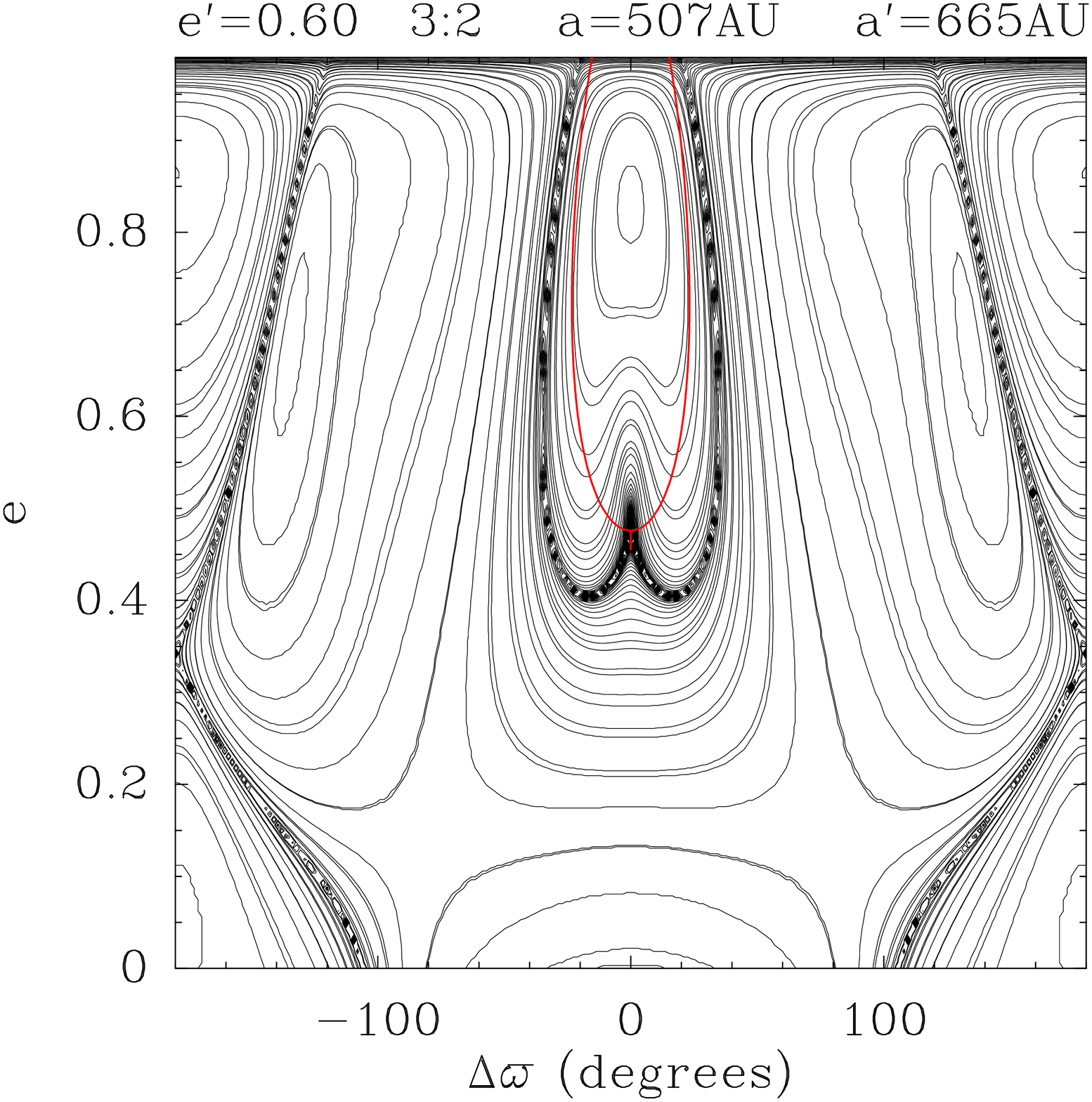}}
%\vspace*{-\jot}\\
%\makebox[\textwidth]{
%\includegraphics[width=0.33\textwidth]{hsec85_bat} \hfil
%\includegraphics[width=0.33\textwidth]{hsec95_bat} \hfil
%\includegraphics[width=0.33\textwidth]{hsec75_bat}}
\caption[]{Phase portraits of resonant Hamiltonian (\ref{hres}) in $(\Delta\varpi,e)$ space after averaging over the orbital motion of P9, for particle having zero resonant amplitude libration (see text), with the same input parameters in Fig.~(\ref{hsec_map}). Each map corresponds to a specific MMR which is specified on top of the plot together with the corresponding semi-major axis value. The red curve has the same meaning as in Fig.~\ref{hsec_map}.}
\label{hsec_res_map}
\end{figure*}
In this section, I explore the dynamics of particles trapped in ($p+q:p$) MMR with P9 ($p$ and $q$ are integers) and perturbed by the giant planets, in a similar semi-analytic way as detailed above. Appendix~\ref{hresapp} describes the way the secular resonant Hamiltonian is obtained. This goes through the introduction of the critical argument of the resonance
\mbox{$\sigma=((p+q)/p)\,\lambda'-(p/q)\,\lambda-\varpi$},
where $\lambda$ and $\lambda'$ are the mean longitudes of the particle and of P9 respectively. As before, I treat the interaction with the known planet in a secular way. After some algebra (see Appendix~\ref{hresapp}), the instantaneous resonant Hamiltonian can be written as
\begin{eqnarray}
\lefteqn{H_\mathrm{res} \; = \; -\frac{GM}{2a}-Gm'\left(\frac{1}{|\vec{r}-\vec{r'}|}-\frac{\vec{r}\cdot\vec{r'}}{r'^3}\right)
-\frac{p+q}{p}\,n'\sqrt{aGM}}&&\nonumber\\
\lefteqn{-\frac{1}{4}\frac{GM}{a(1-e^2)^{3/2}}\sum_{i=1}^4
\frac{m_ia_i^2}{Ma^2}+\dot{\varpi}'\sqrt{aGM}
\left(\frac{p+q}{p}-\sqrt{1-e^2}\right)\:,}&&
\label{hres}
\end{eqnarray}
where $n'=\rd\lambda'/\rd t$ is the mean angular velocity of P9. Due to the resonant configuration, the Hamiltonian cannot be averaged over both orbital motions independently. But the critial angle $\sigma$ is a slow varying variable. 
$H_\mathrm{res}$ can then be averaged for fixed $\sigma$ over the orbital motion of P9. This leaves an autonomous two degrees of freedom Hamiltonian describing the resonant motion.  

When $e'=0$, the resulting Hamiltonian does not depend on $\Delta\varpi$.
In that case, the conjugate action to $\Delta\varpi$
\begin{equation}
N=\sqrt{aGM}\left(\sqrt{1-e^2}-\frac{p+q}{p}\right)\;,
\end{equation}
is a constant of motion, and the planar problem becomes integrable \citep{moo95}. The resonant motion is characterized by libration of $\sigma$ around a secular equilibrium, combined with eccentricity and semi-major axis oscillations. But P9's orbit is eccentric. It can nevertheless be shown \citep{mor93} that the motion is still characterized by coupled libations of $\sigma$, $a$, and $e$. The action $N$ is no longer a constant, and its value is subject to changes on a longer timescale. This slower drift of $N$ can drive the particle towards high eccentricity regime. 
%This mechanism is supposed to be the source of the Falling Evaporating Bodies (FEBs) in the $\beta\:$Pictoris system \citep{beu96,beu07}, and has been invoked as a possible origin of the high eccentricity of the imaged exoplanet Fomalhaut~b \citep{far15}.

The slower drift of $N$ is characterized by the preservation of a new action $J$ \citep{mor93,hen90} that is related to the amplitude of the $\sigma$-libration. A good way to investigate the dynamics is then to consider negligible libration amplitude orbits. 
The preservation of $J$ ensures that an orbit with small initial libration 
amplitude will keep it during the $N$-drift process.
For such orbits, the semi-major axis remains unchanged. After averaging, the resonant Hamiltonian (\ref{hres}) reduces thus to one degree of freedom so that phase portraits can be drawn. This technique was first used by \citet{yosh89} to explore the dynamics of inner resonances with Jupiter, and furthermore by \citet{beu96} to study dynamical routes that may generate star-grazing comets in the $\beta\:$Pictoris system. 
%It was recently applied by \citet{far15} to the Fomalhaut~b case. Subsequent nu%merical simulations \citep{beu00, far15} have demonstrated that particles with %moderate libration amplitudes actually follow the same dynamical paths.

I present in Fig.~\ref{hsec_res_map} phase portraits for a selection of MMRs, all computed still assuming the same parameters for P9, except that I have set P9's semi-major axis to 665\,au instead of 700\,au. This makes the 3:2 MMR to fall at $a=507\,$au, the semi-major axis value quoted by \citet{mal16} for Sedna, who suggest indeed that Sedna could be trapped in that MMR with P9. I have tested many more MMRs. Those presented here have been selected because they fall in the suitable range of semi-major axis, and because they present anti-aligned libration islands at high eccentricity. Some other resonances, like 9:5, 8:5 and 7:5 also present similar libration islands. Particles trapped in these MMRs evolving in the quoted anti-aligned islands could well match the particles depicted by B16 in their simulations.
\section{Discussion}
The high eccentricity, anti-aligned librating particles quoted by B16 as representative for the distant KBOs in relationship with the hypothetical P9 could well correspond to bodies trapped in some of the MMRs listed above. But they could also be non-resonant particles as well, subject to regular secular dynamics as shown in Fig.~\ref{hsec_map}. It is difficult to state which process is dominant. Contrary to resonances, non-resonant configurations have the advantage of not being confined to specific semi-major axis locations with respect to P9. They may therefore concern many more particles. However, the particles in the anti-aligned libration islands move on orbits that cross that of P9. They are thus subject to possible ejection via close encounters, but this may occur after a long time. Numerical simulations by \citet{beu14} applied to the case of Fomalhaut showed that particles moving on orbits that cross that of an eccentric $\sim 10\,\mbox{M}_\oplus$ planet may survive hundreds of Myrs before being ejected. The planet considered in the Fomalhaut simulations had an orbital period of 1000\,yr. With $a'=700\,$au, the hypothetical P9 has a similar mass, but a period of 15,000\,yr. The survival time of any particle crossing its orbit should thus scale similarly and could be comparable to the age of the Solar System.

Figure~\ref{hsec_map} also shows aligned libration islands ($\Delta\varpi=0$) next to anti-aligned ones. Bodies trapped in these islands are potentially longer-lived than anti-aligned ones, as they move in non-crossing regions of phase-space (see the red curves in Fig.~\ref{hsec_map}). However, as pointed out by B16, these islands are located too low in eccentricity to match the observed distant KBOs. Nonetheless, is P9 is real, numerous bodies should be present in these islands, but they are beyond observing capabilities.

Conversely, as suggested by B16, the survival of resonant particles despite crossing orbits is facilitated by the phase protection mechanism. More specifically, the confinement of the critical angle $\sigma$ around an equibrium position prevents the particle from encountering the planet at the time it crosses its orbit \citep{mor93,moo95}. In some cases however, this situation may not last for ever. Due to perturbations by other planets, particles can have large libration amplitudes and be extracted from resonances. B16 note indeed in their simulations particles exhibiting temporary resonant trapping. 

It turns out that any of the two mechanisms outlined here has its own advantages and disadvantages. Stating which one is dominant is difficult, because it should depend on the specific configuration. The constraints on the hypothetical P9 are still too weak to conclude, but I stress here that the anti-aligned librating particles quoted by B16 are not necessarily resonant, but could reside there thanks to regular secular dynamics as well. Hopefully a better statistics on distant KBOs will help refining this analysis in the next future.
%
%\begin{acknowledgements}
%Computations presented in this paper were performed
%using the Froggy platform of the CIMENT infrastructure
%(https://ciment.ujf-grenoble.fr), which is supported by the
%Rh\^one-Alpes region (GRANT CPER07\_13 CIRA), the OSUG@2020 labex
%(reference ANR10 LABX56) and the Equip@Meso project (reference
%ANR-10-EQPX-29-01) of the programme Investissements d'Avenir,
%supervised by the Agence Nationale pour la Recherche.
%
%\end{acknowledgements}
%

%
\Online
\begin{appendix}
\section{The non-resonant secular Hamiltonian}
\label{hsecapp}
I describe here the averaging and expansion process of the non-resonant secular Hamiltonian. In the framework of the restricted three-body problem, the instantaneous interaction Hamiltonian between the particle and P9 reads
\begin{equation}
H=-\frac{GM}{2a}-Gm'\left(\frac{1}{|\vec{r}-\vec{r'}|}-\frac{\vec{r}\cdot
\vec{r'}}{r'^3}\right)\;,
\end{equation}
where $M$ is the mass of the Sun, $a$ is the particle's semi-major axis, and $\vec{r}$ and $\vec{r'}$ are the heliocentric radius vectors of the particle and of P9, respectively. The secular Hamiltonian is obtained taking the time average of $H$ over both orbits independently. This reads
\begin{equation}
H_\mathrm{sec,0}=-\frac{GM}{2a}-\frac{Gm'}{4\pi^2}\int_0^{2\pi}\!\!\int_0^{2\pi}
\left(\frac{1}{|\vec{r}-\vec{r'}|}-\frac{\vec{r}\cdot
\vec{r'}}{r'^3}\right)\,\rd l\,\rd l'\;,
\end{equation}
where $l$ and $l'$ are the mean
anomalies of the particle and P9 respectively. Note that $a$ is now a secular invariant. This a consequence of the averaging process, as the secular Hamiltonian does not depend of the mean longitudes. 

Then, as explained by B16, terms must be added to $H_\mathrm{sec,0}$ to account for the perturbing action of the giant planets. I first add the direct term, so that we have now
\begin{equation}
H_\mathrm{sec,1} = H_\mathrm{sec,0}-\frac{1}{4}\frac{GM}{a}\left(1-e^2\right)^{-3/2}\sum_{i=1}^4\frac{m_ia_i^2}{Ma^2}\;.
\end{equation}
This Hamiltonian is implicitly expressed as a function of the canonically conjugate planar Delaunay orbital elements of the particle, namely
\begin{equation}
Q=\left(\begin{array}{l}\lambda\\
\varpi\end{array}\right)\qquad
P=\left(\begin{array}{l}\sqrt{aGM}\\
\sqrt{aGM}\left(\sqrt{1-e^2}-1\right)\equiv\Phi\end{array}\right)\;.
\label{delaunay}
\end{equation}
As the semi-major axis is a secular invariant, $H_\mathrm{sec,1}$ reduces to one degree of freedom with ($\varpi,\Phi$) as conjugate variable. The constant Keplerian term $-GM/2a$ can also be removed from it. $H_\mathrm{sec,1}$ is nevetheless time-dependent as $\varpi'$ is not constant. I thus perform a canonical transformation, changing $\varpi$ to $\Delta\varpi$. To do this, I use the generating function 
$S=-\Phi\varpi=-\Phi(\Delta\varpi+\varpi')$. The new momentum conjugate to $\Delta\varpi$ is still $\Phi$, and the new autonomous Hamiltonian reads
\begin{equation}
H_\mathrm{sec}=H_\mathrm{sec,1}+\frac{\partial S}{\partial t}=
H_\mathrm{sec,1}+\Phi\dot{\varpi}'\qquad,
\end{equation}
which is Eq.~(\ref{htot}) from the text.

Unfortunately,
the above expression cannot be computed in closed form, so that to get a closed form expression, expansion of the disturbing function is required before averaging. If the particle's orbit lies inside P9's one (which is the situation supposed here), then the instantaneous Hamiltonian can be expanded using Legendre polynomials $P_k$ ($k\ge0$):
 \begin{equation}
H = -\frac{GM}{2a}-\frac{Gm'}{r'}\left(1+ \sum_{k=2}^{+\infty}\left(\frac{r}{r'}\right)^kP_k(\cos\beta)\right)\;,
\label{legexp}
\end{equation}
where $\beta$ is the angle between radius vectors $\vec{r}$ and $\vec{r'}$. Each term of the sum can then be averaged over both orbital motions independently. The result reads
\begin{eqnarray}
H_\mathrm{sec,2} & = & -\frac{1}{4}\frac{GM}{a}\left(1-e^2\right)^{-3/2}\sum_{i=1}^4
\frac{m_ia_i^2}{Ma^2}\nonumber\\
&&\qquad+\Phi\dot{\varpi}'-\frac{Gm'}{a'}\left(
1+\sum_{k=2}^{+\infty}h_k\right)\;.
\end{eqnarray}
where
\begin{equation}
h_k=\frac{1}{4\pi^2}\int_0^{2\pi}\!\!\int_0^{2\pi}\left(\frac{r}{a}\right)^k
\left(\frac{r'}{a'}\right)^{-k-1}P_k(\cos\beta)\,\rd l\,\rd l'\:.
\end{equation}
Each $h_k$ is proportional to $(a/a')^k$ and can be expressed in closed form as a function of the orbital elements of both orbits with growing complexity, making use of so-called Hansen coefficients. Details about this process are given in \citet{las10} with explicit expressions of the $h_k$'s up to $k=10$. In the case of coplanar orbits, apart from being proportional to $(a/a')^k$, the $h_k$'s are function of $e$, $e'$ and of $\Delta\varpi=\varpi'-\varpi$ only, where $\varpi$ and $\varpi'$ are the longitudes of periastron of the particle and of P9 respectively. As an example, the first ones read
\begin{eqnarray}
h_2 & = & \frac{1}{4}\left(\frac{a}{a'}\right)^2\frac{(3/2)e^2+1}{(1-e'^2)^{3/2}}\;,\\
h_3 & = & -\frac{3}{16}\left(\frac{a}{a'}\right)^3e\,e'\frac{(15/4)e^2+5}{
(1-e'^2)^{5/2}}\,\cos\left(\Delta\varpi\right)\;,\\
h_4 & = & \left(\frac{a}{a'}\right)^4\frac{1}{(1-e'^2)^{7/2}}\left[
\frac{15}{256}\left(\frac{21}{2}e^2+1\right)e^2\,e'^2\,\cos\left(2\Delta\varpi\right)
\right.\nonumber\\
&&\left.
+\frac{9}{64}\left(\frac{15}{8}e^4+5e^2+1\right)\left(1+\frac{3}{2}e'^2\right)\right]\;.
\end{eqnarray}
\section{The resonant secular Hamiltonian}
\label{hresapp}
I consider now a situation where the particle is trapped in \mbox{($p+q:p$)} MMR with P9 and perturbed by the giant planets. As above I treat the interaction with the giant planet in a secular way, so that my starting instantaneous Hamiltonian reads
\begin{eqnarray}
H_\mathrm{res,0} & = & -\frac{1}{4}\frac{GM}{a}\left(1-e^2\right)^{-3/2}\sum_{i=1}^4
\frac{m_ia_i^2}{Ma^2}\nonumber\\
&&\qquad-\frac{GM}{2a}-Gm'\left(\frac{1}{|\vec{r}-\vec{r'}|}-\frac{\vec{r}\cdot
\vec{r'}}{r'^3}\right)\;,
\end{eqnarray}
This Hamiltonian is still implicitly expressed as a function of the Delaunay elements (\ref{delaunay}), but now it does not immediately reduce to one degree of freedom. Due to the resonance, the semi-major axis can indeed have secular variations. $H_\mathrm{res,0}$ cannot thus be averaged over both orbits independently. I introduce now the critical argument of the resonance
\begin{equation}
\sigma=\frac{p+q}{p}\,\lambda'-\frac{p}{q}\,\lambda-\varpi\;,
\end{equation}
and the new coordinate vector $Q_1=(\sigma,\Delta\varpi)$. I have $Q_1=AQ+B(t)$, where
\begin{equation}
A=\left(\begin{array}{ll} -p/q & -1\\0 & 1\end{array}\right)
\quad\mbox{and}\quad B(t)=\left(\begin{array}{l} -(p+q)/q\,\lambda'\\
-\varpi'\end{array}\right)
\end{equation}
Considering now the generating function 
\mbox{$S(P,Q_1)=-^tPQ=-^tPA^{-1}(Q_1-B(t))$}, I perform a canonical transformation with $Q_1$ as new coordinate vector, and $P_1=\,\mbox{}^tA^{-1}P$ as new momenta vector: 
\begin{equation}
Q_1=\left(\begin{array}{l}\sigma\\
\Delta\varpi\end{array}\right)\;
P_1=\left(\begin{array}{l}-(q/p)\sqrt{aGM}\\
\sqrt{aGM}\left(\sqrt{1-e^2}-(p+q)/p\right)\end{array}\right)\;.
\end{equation}
The new Hamiltonian then reads
\begin{eqnarray}
\lefteqn{H_\mathrm{res} \;= \; H_\mathrm{res,0}+\frac{\partial S}{\partial t}\;=\;H_0-\,^tPA^{-1}\frac{\partial B}{\partial t}\;=\;
H_\mathrm{res,0}-\,^tP_1\frac{\partial B}{\partial t}}&&\nonumber\\
\lefteqn{=H_\mathrm{res,0}+\dot{\varpi}'\sqrt{aGM}
\left(\frac{p+q}{p}-\sqrt{1-e^2}\right)-\frac{p+q}{p}\,n'\sqrt{aGM}\:,}&&
\end{eqnarray}
where $n'=\rd\lambda'/\rd t$ is the mean angular velocity of the perturbing planet. This corresponds to Eq.~(\ref{hres}) from the text.
\end{appendix}

\begin{thebibliography}{}
%
\bibitem[Aarseth(1999)]{aar99} Aarseth S.J., 1999, Celest. Mech. 73, 127
%
\bibitem[Batygin \&\ Brown(2016)]{bat16} Batygin, K., \&\ Brown, M.E. 2016, ApJ, 151, 22 (B16)
%
\bibitem[Beust \&\ Morbidelli(1996)]{beu96} Beust, H., \&\ Morbidelli, A. 1996, 
Icarus, 120, 358
%
%\bibitem[Beust \&\ Morbidelli(2000)]{beu00} Beust, H., \&\ Morbidelli, A. 2000, Icarus, 143, 170
%
%\bibitem[Beust \&\ Valiron(2007)]{beu07} Beust, H., \&\ Valiron, P. 2007, A\&A, 466, 201
%
\bibitem[Beust et al.(2014)]{beu14} Beust, H., Augereau, J.-C., Bonsor, A., et al. 2014, A\&A, 561, A43
%
\bibitem[Cowan et al.(2016)]{cow16} Cowan, N.B., Holder, G., \&\ Kaib, N.A., 2016, arXiv:1602.05963v1
%
%\bibitem[Faramaz et al.(2015)]{far15} Faramaz, V., Beust, H., Augereau, J.-C., Kalas, P., \&\ Graham, J.R. 2005, A\&A, 573, A87
%
\bibitem[Fienga et al.(2016)]{fie16} Fienga, A., Laskar, J., Manche, H., \&\ Gastineau, M. 2016, A\&A, submitted, arXiv:1602.06116v1
%
\bibitem[de la Fuente Marcos \&\ de la Fuente Marcos(2014)]{fue14}
de la Fuente Marcos, C., \&\ de la Fuente Marcos, R., 2014, Ap\&SS, 352, 409
%
\bibitem[Henrard(1990)]{hen90} Henrard, J. 1990, Celest. Mech., 49, 43  
%
\bibitem[Laskar \&\ Bou\'e(2010)]{las10} Laskar, J., \&\ Bou\'e, G. 2010, A\&A 522
%
\bibitem[Malhotra et al.(2016)]{mal16} Malhotra, M., Volk, K., \&\ Wang, X. 2016, submitted, arXiv:1603.02196v2
%
\bibitem[Moons \&\ Morbidelli(1995)]{moo95} Moons, M., \&\ Morbidelli, A. 1995, Icarus 114, 33
%
\bibitem[Morbidelli \&\ Moons(1993)]{mor93} Morbidelli, A., \&\ Moons, M. 1993, Icarus 102, 316
%
\bibitem[Trujillo \&\ Sheppard(2014)]{tru14} Trujillo, C.A., \&\ Sheppard, S.S. 2014, Nature, 507, 471
%
\bibitem[Yoshikawa(1989)]{yosh89} Yoshikawa, M. 1989, A\&A, 213, 436
%
\end{thebibliography}
\end{document}